\documentclass[aps,pra,twocolumn,nofootinbib,showkeys]{revtex4-2}
\makeatletter
\def\l@subsubsection#1#2{}
\makeatother

\usepackage[hyperref]{xcolor}

	\definecolor{fsu-blue}{cmyk}{1.0,0.7,0.1,0.5}			
	\definecolor{fsu-gold}{cmyk}{0.2,0.25,0.6,0.25}			
	\definecolor{fsu-violet}{cmyk}{0.8,0.85,0.0,0.0}		
	\definecolor{fsu-purple}{cmyk}{0.5,1.0,0.0,0.1}			
	\definecolor{fsu-magenta}{cmyk}{0.15,1.00,0.25,0.15}	
	\definecolor{fsu-red}{cmyk}{0.0,0.95,0.8,0.15}			
	\definecolor{fsu-orange}{cmyk}{0.1,0.7,1.0,0.0}			
	\definecolor{fsu-yellow}{cmyk}{0.05,0.4,1.0,0.1}		
	\definecolor{fsu-light-green}{cmyk}{0.7,0.1,1.0,0.05}	
	\definecolor{fsu-green}{cmyk}{0.9,0.3,1.0,0.1}			
	\definecolor{fsu-turquoise}{cmyk}{1.0,0.25,0.4,0.1}		
	\definecolor{fsu-light-blue}{cmyk}{0.7,0.2,0.0,0.2}		

    \definecolor{py-maroon}{rgb}{0.5,0,0}
    \definecolor{py-navy}{rgb}{0,0,0.5}
    \definecolor{py-green}{rgb}{0.0,0.5,0.0}
    \definecolor{py-olivedrab}{rgb}{0.4196,0.5569,0.1373}
    \definecolor{py-dodgerblue}{rgb}{0.1176,0.56471,1.0}
    \definecolor{py-crimson}{rgb}{0.8627,0.0784,0.2353}
    \definecolor{py-gray}{rgb}{0.5,0.5,0.5}

\usepackage[utf8]{inputenc}

\usepackage[
	colorlinks,
	pdfpagelabels,
	breaklinks,
	pdfstartview=FitH,
	bookmarksopen=true,
	bookmarksnumbered=true,
	bookmarksopenlevel=2,
	plainpages=false,
	hypertexnames=false,
	citecolor=fsu-gold,
	linkcolor=fsu-gold,
	urlcolor=fsu-gold,
	pdftitle={The massive Thirring / sine-Gordon model with non-zero current density},				
	pdfauthor={Oevermann and Cohen},	
	pdfkeywords={}					
]{hyperref}

\usepackage[acronym]{glossaries-extra}
	\glssetcategoryattribute{acronym}{nohyperfirst}{true}
	\setabbreviationstyle[acronym]{long-short}
	
	\usepackage{amsmath}
	\usepackage{amssymb}
	\usepackage{bm}
	\usepackage{bbm}
	\usepackage{slashed}
	\usepackage{cancel}

	\usepackage{tensor}
	\usepackage{upgreek} 
	\usepackage{amsbsy}
	\usepackage{mathtools}
	\usepackage{lipsum}
	\usepackage{multirow} 
	\usepackage{mathrsfs}

	\usepackage{comment}
	\usepackage{soul}
    \usepackage{graphicx}
    \usepackage[caption=false]{subfig}
    \usepackage{placeins}
	\usepackage{booktabs}
	\usepackage{tabularx}
	\usepackage{pifont}
	\usepackage{orcidlink}
    \usepackage{enumitem}
    \usepackage{tikz}
\usepackage{flushend}

\usepackage[capitalise,sort,compress]{cleveref}
\crefname{equation}{eq.}{eqs.}
\Crefname{equation}{Eq.}{Eqs.}

\crefname{section}{sec.}{secs.}
\Crefname{section}{Sec.}{Secs.}

\crefname{subsection}{sec.}{secs.}
\Crefname{subsection}{Sec.}{Secs.}

\crefname{figure}{fig.}{figs.}
\Crefname{figure}{Fig.}{Figs.}

\crefname{table}{tab.}{tabs.}
\Crefname{table}{Tab.}{Tabs.}

\allowdisplaybreaks

\usepackage[acronym]{glossaries-extra}
\glssetcategoryattribute{acronym}{nohyperfirst}{true}
\setabbreviationstyle[acronym]{long-short}

\newacronym{wrt}{w.r.t.}{with respect to}
\newacronym{rhs}{r.h.s.}{right hand side}
\newacronym{lhs}{l.h.s.}{left hand side}

\newacronym{mtm}{MTM}{massive Thirring model}
\newacronym{sgm}{SGM}{sine-Gordon model}
\newacronym{eos}{EoS}{equation of state}

\newcommand{\Reff}{ref.~}

\newcommand{\App}{app.~}
\newcommand{\Eq}{eq.~}

\newcommand{\ie}{\textit{i.e.}}

\newcommand{\cf}{\textit{c.f.}}

\newcommand{\ii}{\mathrm{i}}
\newcommand{\ee}{\mathrm{e}}
\newcommand{\dd}{\mathrm{d}}

\begin{document}


\title{
	The massive Thirring / sine-Gordon model with non-zero current density
}
    
\author{Eric Oevermann \orcidlink{0009-0000-9431-3670}}
	\email{eric.oevermann@uni-jena.de}
	\affiliation{
		Institute for Theoretical Physics, 
        Friedrich Schiller University Jena, 
        07743 Jena, Germany
	}

\author{Thomas D. Cohen \orcidlink{0000-0002-4370-3297}}
	\email{cohen@umd.edu}
	\affiliation{
		Department of Physics,
        University of Maryland, 
        College Park, MD 20742 USA
	}

\date{\today}

\begin{abstract}
   
This paper determines the zero-temperature equation of state for the massive Thirring / sine-Gordon model.  This demonstrates recently derived  model‑independent upper and lower bounds on the zero‑temperature equation of state with fixed number density from systems with a non‑zero current density.  That approach is potentially valuable as Monte Carlo calculations with a current density avoid the sign problem in the Euclidean formulation.  An advantage to illustrating these bounds in the massive Thirring / sine-Gordon model is that the relevant calculations with both a number density and a current density can be done using a Bethe ansatz.  For this model, optimal bounds constrain the energy density as a function of number density by a factor of two from above and below at high densities for all choices of couplings. The lower bound becomes exact at low densities, while the upper bound approaches the worst constraint of a factor of $ 4.90 $. 
\end{abstract}

\keywords{Equation of state, massive Thirring model / sine-Gordon model, Bethe ansatz}

\maketitle



\section{Introduction}

\raggedbottom
Understanding the \gls{eos} of QCD in extreme environments with nonzero temperatures or baryon number density or external magnetic fields has been of interest to the community for a significant period and remains of interest \cite{Philipsen_2013,Ding:2015ona,SP0,Monnai_2021,NS1,NS2,PhysRevC.80.015203,RevModPhys.88.025001,Bali_2014,Adhikari:2024bfa}.  In part, this is to understand environments that arise in experiments or astronomical observations.  But there is also substantial interest from a purely theoretical perspective, even in regimes that do not appear to be immediately relevant to current or foreseeable experimental measurements.  In this context, it is noteworthy that one class of extreme environment has remained largely unstudied.  While environments in which the number density of a conserved $U(1)$ current (such as baryon number in QCD) is nonzero have been of wide interest, environments in which the current density is nonzero have not.  For the case of QCD with zero baryon density and nonzero baryon current density, one can envision such an environment as corresponding to baryons flowing in one direction and and antibaryons flowing in the opposite direction. 

Of course, it is not surprising that QCD with a nonzero baryon current density is of less interest than QCD with a nonzero baryon number density: the latter is a familiar situation and is of  direct relevance to some experimental situations.  However, while the former situation is much less familiar and does not directly arise in situations of experimental interest,  it ought to be of some theoretical interest on its own.  Moreover, these properties (at zero number density) should be accessible in lattice simulations without encountering an intractable sign problem.

Recently, a new reason has emerged to study the properties of field theories such as QCD with nonzero current densities, at least at zero temperature: it is the basis of a  novel, model-independent approach that exploits relativity \cite{Cohen:2025ahp} and constrains the \gls{eos}.  In particular, it provides upper and lower bounds for the energy density as a function of number density.    This is significant since direct calculations of the \gls{eos}  are intractable using standard lattice methods due to a notorious sign problem \cite{SP0,SP1,SP2}.  In contrast, calculations for systems with a nonzero current density (and zero number density) might evade the sign problem.  Thus, this approach provides a possible  method to constrain the zero temperature \gls{eos} while evading the sign problem.

This is important since the properties of nuclear matter at nonzero density and zero temperature are key issues in both nuclear physics and in astrophysical studies of neutron stars.
The sign problem prevents direct calculations. Attempts at analytic continuation from imaginary to real potentials are limited  to large temperatures \cite{D_Elia_2003}.  For instance, they fail at zero temperature due to the Silverblaze phenomenon \cite{Cohen:2003kd,cohen2026silverblazeproblemqcd}.  While non-perturbative functional approaches \cite{PhysRevD.102.034027} might allow in the future reliable and converged calculations with nonzero number density, they are not yet advanced enough to determine the entire QCD phase diagram. 
In the meantime, one might hope to at least constrain the zero temperature \gls{eos}. 

In this paper, we compute the energy  density as a function of current (the analog of current density in ${1+1}$ dimensions) at zero number density within the \gls{mtm} or equivalently its dual, the \gls{sgm} at various values of the coupling constant of the theory.  As noted above, this quantity should be of at least modest theoretical interest in its own right.  It is also the key  input into an examination of the usefulness of the bounds derived in \Reff\cite{Cohen:2025ahp}.  This paper analyzes  these bounds by comparing the at finite current calculations with analogous calculations for finite number density.
It is the first illustration of these bounds using a nontrivial quantum field theory.  

Note that in order to illustrate how well the bounds work, it is necessary to be able to compute properties of systems with non-zero number densities as well as with non-zero current densities.  This is not possible for most nontrivial quantum field theories, since computations with non-zero number densities are not viable.  However, for \gls{mtm}/\gls{sgm} computations for non-zero number densities are viable using Bethe ansatz techniques, these calculations involve nothing more numerically challenging than solving one-dimensional integral equations that were determined by Haldane more than four decades ago \cite{Haldane1982}.  In our calculations for nonzero currents, we generalized Haldane's method to this new situation.  While this situation should be tractable by Euclidean space lattice Monte Carlo methods, the Bethe ansatz approach is numerically cleaner, particularly for zero temperature (which would require an extrapolation in a lattice simulation).
Thus, the extent of how constraining these bounds are can be easily checked within the \gls{mtm}/\gls{sgm}. 
    
    We begin by reviewing some of the model's properties at non-zero number density. 
    In \cref{sec:formalism}, we derive an integral equation for the density of states for a theory with a non-zero current density. 
    It allows us to compute the bounds for the \gls{mtm}, and we explain our results.

\section{Massive Thirring / Sine-Gordon model}
\label{sec:model}

  One of the most remarkable discoveries involving field theory in the 1970s, was that the \gls{mtm}---a bosonic theory---and the \gls{sgm}---a fermionic theory---were dual to each other, {\it i.e.} were essentially the same theory cast into very different languages  \cite{Coleman:1974bu,Mandelstam:1975hb}.   The 
 Lagrangians of the two theories
        \begin{subequations}
            \begin{align}
                \mathcal{L}_\text{MTM} = \, & \bar{\psi} \, ( \ii \, \gamma^\mu \, \partial_\mu - M_0 ) \, \psi - \frac{g}{2} \, ( \bar{\psi} \, \gamma^\mu \, \psi )^2 \, ,
                \\
                \mathcal{L}_\text{SGM} = \, & \frac{1}{2} \, \partial^\mu \phi \, \partial_\mu \phi + \frac{m_0^2}{\beta^2} \, \cos( \beta \phi ) \, ,
            \end{align}
        \end{subequations}
    are clearly radically different.  But the predictions of the two theories are identical
 if one identifies the couplings \cite{Coleman:1974bu,Mandelstam:1975hb} according to
        \begin{align}
            & 4\pi/\beta^2 = 1 + g/\pi > 0 \, ,
            && - \pi / 2  < g \, ,
        \end{align}
    and analogous quantities appropriately.  Note that with this identification the duality relates the weakly coupled \gls{sgm} to the strongly coupled \gls{mtm}. Remarkably,x the conserved topological current in the \gls{sgm} plays an identical role as the $U(1)$ Noether current in the \gls{mtm} \cite{Rajaraman:1982is},
        \begin{align}
            & \mathcal{J}_{\mathrm{SGM}}^\mu =  \beta / (2\pi) \, \epsilon^{\mu\nu} \, \partial_\nu \phi \, ,
            && \mathcal{J}_{\mathrm{MTM}}^\mu =  \bar{\psi} \, \gamma^\mu \, \psi \, .
        \end{align}
    The topological winding number in the \gls{sgm} is identified with  particle number ${ N = \int \dd x \, \bar{\psi} \gamma^\mu \psi }$ in the  \gls{mtm}.  
    
    Quantum mechanically the Hilbert space of the theory is divided into sectors of fixed  integer $N$. The lightest state of ${N=1}$ is a topological soliton in the \gls{sgm} version of the theory and the fundamental fermion of the \gls{mtm}. Upon renormalization, the bare fermion mass $ M_0 $ obtains a multiplicative factor. 
    The renormalized mass $ M_\mathrm{r} $ can be identified with the soliton mass $ m_\mathrm{s} $ of the \gls{sgm}: ${ M_\mathrm{r} = m_\mathrm{s} }$. 
    Similarly, the fermion-antifermion bound states of the \gls{mtm} can be identified with the soliton-antisoliton bound states of the \gls{sgm}. 
    The lowest-lying bound state of mass $ m_\mathrm{b} $, also called ``breather'' \cite{Haldane1982}, defines an important length scale and can be identified with the elementary boson in the \gls{sgm} \cite{Cohen:1990zz,Rajaraman:1982is}.
    Before addressing finite density, we introduce our parameterization of the couplings via $ \kappa $ \cite{Cohen:1988fu},
        \begin{align}\label{eq:kappa_definition}
            \kappa \coloneq \frac{\beta^2/(8\pi)}{1-\beta^2/(8\pi)} = \frac{1}{1+2g/\pi} \, ,
   \end{align}
    and distinguish the regimes of weakly (strongly) coupled \gls{sgm} (\gls{mtm}) for ${ \kappa \ll 1 }$, strongly (weakly) coupled \gls{sgm} (\gls{mtm})  for ${ \kappa \simeq 1 }$, and repulsive solitons for ${ \kappa > 1 }$. 
    In this notation, the breather and soliton mass (in the attractive case) are related as follows \cite{Haldane1982}
        \begin{align}\label{eq:relation_mb_ms}
            m_\mathrm{b} = 2 \, \sin( \pi / 2 \, \kappa ) \, m_\mathrm{s} \overset{\kappa \rightarrow 0 }{=} \pi \, \kappa \, m_\mathrm{s} \, .
        \end{align}

\subsection*{Fixed number density}

    To facilitate the relating of our results, we review some of the known physics of the \gls{mtm} in different density regimes. 
    Bergknoff and Thacker used a Bethe ansatz to study the spectrum and vacuum properties of the model \cite{Bergknoff:1978bm,Thacker:1980ei}. 
    Haldane extended these results to non-zero number densities \cite{Haldane1982}. 
    Utilizing the Bethe ansatz technique, one can derive an integral equation for the (renormalized) density of states $ \rho ( \alpha ) $ as a function of rapidity $ \alpha $, 
        \begin{align}\label{eq:integral_eq_Haldane}
            & \rho ( \alpha ) = ( 2 \pi )^{-1} \, \cosh ( \alpha ) + \int_{ -\alpha_{\mathrm{F}} }^{\alpha_{\mathrm{F}}} \dd \alpha' R( \alpha - \alpha' ) \, \rho ( \alpha ) \, ,
        \end{align}
    where $ \alpha_{\mathrm{F}} $ is the Fermi-rapidity. 
    The kernel $ R $ is given by 
        \begin{equation}
            R(\alpha) = \frac{1}{2 \pi} \int_{- \infty}^{\infty} \dd y \, 
            e^{\ii \alpha y} \,  \frac{1}{2} \, \left (1-\frac{ {\rm tanh}(\frac 12 \pi y)  }{{\rm tanh}(\frac 12 \pi \kappa y)} \right)\, ,
        \end{equation}
    where $\kappa$ is the effective coupling constant given in \cref{eq:kappa_definition}. 
    From the density of states, quantities such as the number $ n $ and energy density $ \epsilon $ can be obtained. 
    We go in more detail about this method, deriving our new result for non-zero current densities below in \cref{eq:final_density_of_states_integral_equation_kappa} and \App\labelcref{app:Bethe_an_int_eq}. 

    We distinguish between high- and low-density regimes \cite{Haldane1982}. 
    The crossover is characterized by the length scale set by the breather mass $ m_\mathrm{b} $ and not $ m_\mathrm{s} $, \cf{} \cref{eq:relation_mb_ms}. 
    Haldane calls this length scale the ``breather length'' and it becomes the effective size of bare solitons dressed by virtual breather fluctuations. 
    At low densities, the system behaves like a non-interacting Fermi gas. 
    We explicitly note the interaction energy per particle \cite[Eq.~(5.2)]{Haldane1982},
        \begin{align}\label{eq:Fermi_gas_regime}
            & \frac{ \epsilon }{ n } - m_\mathrm{s} = \frac{1}{6} \, \frac{\pi^2 \, n^2 }{ m_\mathrm{s} } \, .
        \end{align}
    to compare with the numerical solution of \cref{eq:integral_eq_Haldane} below. 
    At high densities, where $ n / m_\mathrm{b} \gg 1 $, the model behaves just as the massless Thirring model and \cite[Eq.~(5.22)]{Haldane1982}
        \begin{align}\label{eq:high_desity_limit}
            & \frac{ \epsilon }{ n } - m_\mathrm{s} = 4 \, \Big( 1 + \frac{1}{\kappa} \Big) \, n - m_\mathrm{s} \, .
        \end{align}
    A crystalline or mean-field description becomes valid in the limit of ${ \kappa \ll 1 }$, the weakly coupled \gls{sgm}, in the regime of low densities. 
    However, mean-field theory fails to describe the interaction energy at very low densities, requiring the Fermi gas description of \cref{eq:Fermi_gas_regime}, which is explained in detail in \Reff\cite{Cohen:1988fu}. 
    The regimes are shown in \cref{fig:results_weak_strong_coupling}.
    In the mean-field regime, the interaction energy is described by \cite[Eq.~(6.5)]{Haldane1982}
        \begin{align}\label{eq:mean_field_regime}
            & \frac{ \epsilon }{ n } - m_\mathrm{s} = 4 \, \ee^{-m_\mathrm{b}/n} \, m_\mathrm{s} \, .
        \end{align}
    In a QCD language, the interaction is mediated by a single meson exchange.
    At low densities, the dominant contribution to the energy density stems simply from the mass of the individual solitons, which is correctly described by mean field theory. 
    Looking at the interaction energy, which is of more interest from the point of view of nuclear physics, reveals the crossover from a Fermi gas to a low-density mean-field crystalline description.

\section{Fixed current} \label{sec:formalism}

    We seek expressions for the energy-momentum tensor $ T^{\mu\nu} $, \cf{} \cref{eq:bound_eqs}, that minimize the energy density $ \epsilon $ under the constraint of having a non-zero current  $ j $---the analog of current density for this ${(1+1)}$-dimensional system.   
    Many of the details for the construction of this state are presented in \App\labelcref{app:current_dens_ground_state}.
    Ultimately, all quantities of interest can be derived from the density of states $ \rho(\alpha) $, considered as a function of rapidity $ \alpha $. 
    It satisfies the integral equation
        \begin{align}\label{eq:final_density_of_states_integral_equation_kappa}
            & \rho (\alpha) = \frac{\cosh ( \alpha )}{2\pi} +
            \int_{\alpha_1}^{\alpha_2} \dd \alpha' \int_{-\infty}^{\infty} \frac{\dd y}{2\pi} \, \rho (\alpha') 
            \\
            & \quad \times \left[
                e^{\ii y (\alpha-\alpha')} \, \frac{ \bar{ K } ( y ) }{ 1 + \bar{ K } ( y ) } - e^{\ii y (-\alpha-\alpha')} \, \frac{ \bar{ L } ( y ) }{ 1 + \bar{ K } ( y ) }
            \right] \, , \notag 
        \end{align}
    where the Kernel functions $ \bar{ K } $ and $ \bar{ L } $ and all details of the derivation are provided in \App\labelcref{app:Bethe_an_int_eq}. 
    The way in which the integration limits ${ \alpha_1, \alpha_2 \in \mathbb{R} }$ must be chosen is explained in \App\labelcref{app:current_dens_ground_state}. 
    After determining $ \rho(\alpha) $, we compute
        \begin{subequations}\label{eq:j_and_T}
            \begin{align}
                j & = 2 \int_{\alpha_1}^{\alpha_2} \dd \alpha \, n ( \alpha)  \, \tanh \alpha \, , 
                \\
                T^{\mu\nu} & = \begin{cases}
                    \displaystyle 2 \int_{\alpha_1}^{\alpha_2} \frac{ \dd \alpha \, n ( \alpha)  \, p^\mu \, p^\nu }{ m_\mathrm{s} \cosh \alpha } \, , & \mu = \nu \, ,
                    \\
                    0 \, , & \mu \neq \nu \, ,
                \end{cases} \label{eq:energy_momentum_tensor}
            \end{align}
        \end{subequations}
    where the factor two counts anti- and particles, and
        \begin{align}\label{eq:n_and_momentum}
            & n ( \alpha) = m_\mathrm{s} \, \rho (\alpha) \, , 
            & p^{\mu} = m_\mathrm{s} \, ( \cosh{ \alpha }, \sinh{ \alpha } )^{\mu} \, .
        \end{align}
    Appendix \labelcref{app:numerical_setup} describes the details of the numerical setup to obtain data pairs ${ ( j, T^{\mu\nu} ) }$. 
    Using \cref{eq:bound_eqs} and applying boosts ${ \beta \in [0, 1] }$ yields upper and lower bounds on the \gls{eos} for a range of number densities. 
    Repeating the calculation for multiple current densities leads to a family of curves, the envelope of which defines the optimal bounds that are calculable from this approach. 
\section{General bounds on the equation of state}
\label{sec:bounds}

    As discussed in the introduction, one of the motivations for studying this model at fixed current is to illustrate the utility of bounds relating the \gls{eos} as a function of density to results for fixed current density.  These bounds are reviewed in this section.
    
    Assume that the energy-momentum tensor $ T_{\mu\nu} $ of a system with a non-zero current density $ j $ and zero number density is known. 
    It was shown in \Reff\cite{Cohen:2025ahp} that this allows one to infer upper (u) and lower (l) bounds on the energy density  $ \epsilon $ as a function of number density $ n $ via Lorentz boosts with a boost parameter $ \beta $:
        \begin{subequations}\label{eq:bound_eqs}
            \begin{align}
                & \epsilon ( n^\mathrm{u} ) \leq \frac{T_{tt} + \beta^2 \, T_{xx}}{1 - \beta^2} \, , \label{eq:upper_bound}
                \\
                & \epsilon ( n^\mathrm{l} ) \geq \frac{ 1 - \beta^2}{1 + \beta^2 \, \frac{p( n^\mathrm{l} )}{\epsilon(n^\mathrm{l} )}} \, T_{tt}
                \geq \frac{ 1 - \beta^2}{1 + \beta^2 } \, T_{tt} \, , \label{eq:lower_bound}
                \\
                \mathrm{with} \quad & n^\mathrm{u} \coloneq \frac{ \beta \, j }{ \sqrt{ 1 - \beta^2 } } \, , \qquad
                n^\mathrm{l} \coloneq \frac{ \sqrt{ 1 - \beta^2 } \, j }{ \beta } \, . \label{eq:def_densities_l_and_u}
            \end{align}
        \end{subequations}
    The lower bound requires knowledge of the ratio ${ p / \epsilon \leq 1 }$, which is usually unknown but constrained by causality, and fixed in any given proposed model for the \gls{eos}.
    The bound is obtained by boosting a system with a zero current and non-zero number density,
        \begin{align}\label{eq:derivation_lower_bound}
            & \begin{pmatrix}
                n^\mathrm{l} \\ 0
            \end{pmatrix}
            \xrightarrow{\beta}
            \begin{pmatrix}
                n \\ j
            \end{pmatrix} \, ,
            & T_{tt} ( 0, j ) \leq T_{tt}^\beta ( n, j) \, .
        \end{align}
    The upper bound is derived by boosting a system of zero number and non-zero current density, 
        \begin{align}\label{eq:derivation_upper_bound}
            & \begin{pmatrix}
                0 \\ j
            \end{pmatrix}
            \xrightarrow{\beta}
            \begin{pmatrix}
                n^\mathrm{u} \\ j^\mathrm{u}
            \end{pmatrix} \, ,
            & \epsilon ( n^\mathrm{u} ) \leq T_{tt}^\beta ( n^\mathrm{u}, j^\mathrm{u}) \, .
        \end{align}

    We work exclusively in Minkowski space, calculating the bounds within the \gls{mtm}. 
    However, a Lagrange multiplier for a finite current density in Minkowski spacetime, when continued to the Euclidean domain, is equivalent to an imaginary chemical potential in Minkowski spacetime, when translated to the Euclidean domain. 
    Therefore, there is no sign problem in practical lattice calculations.

\section{Numerical results}
\label{sec:results}

\begin{figure*}[htb]               
          \centering
          \subfloat[$\kappa = 0.001.$\label{fig:kappa_0_001}]%
          {%
            \includegraphics[width=.49\linewidth]{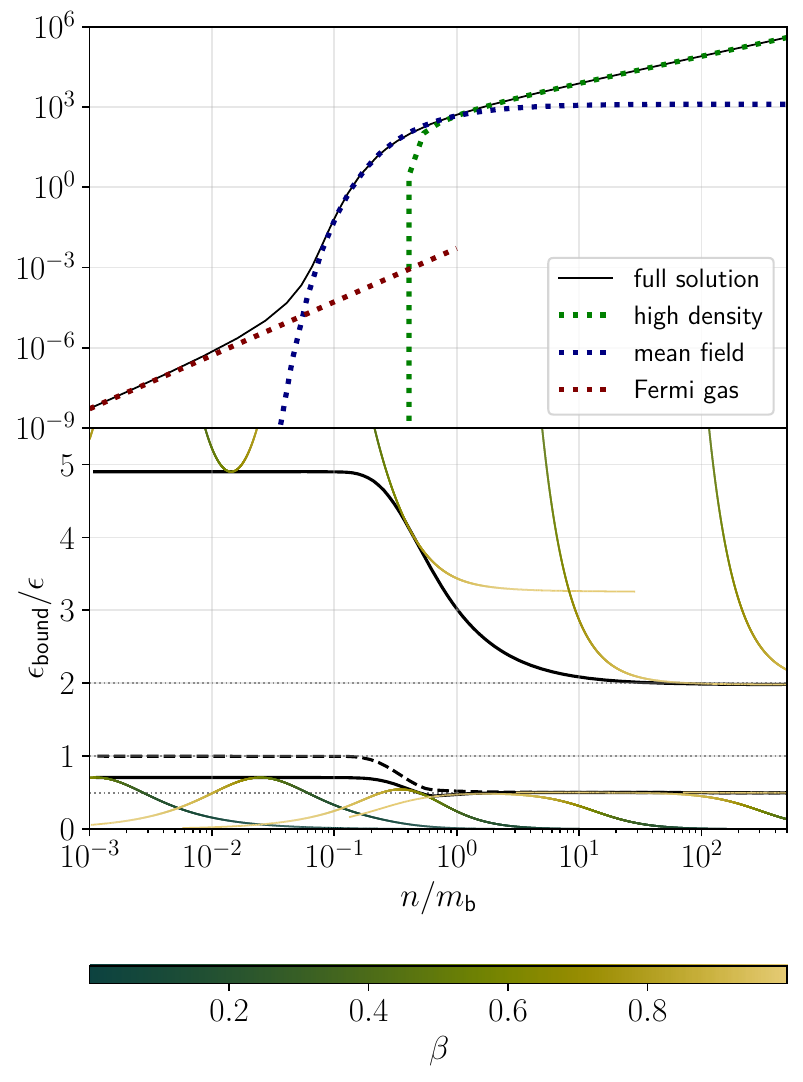}%
          }
          \subfloat[$\kappa = 0.99.$\label{fig:kappa_0_99}]%
          {%
            \includegraphics[width=.49\linewidth]{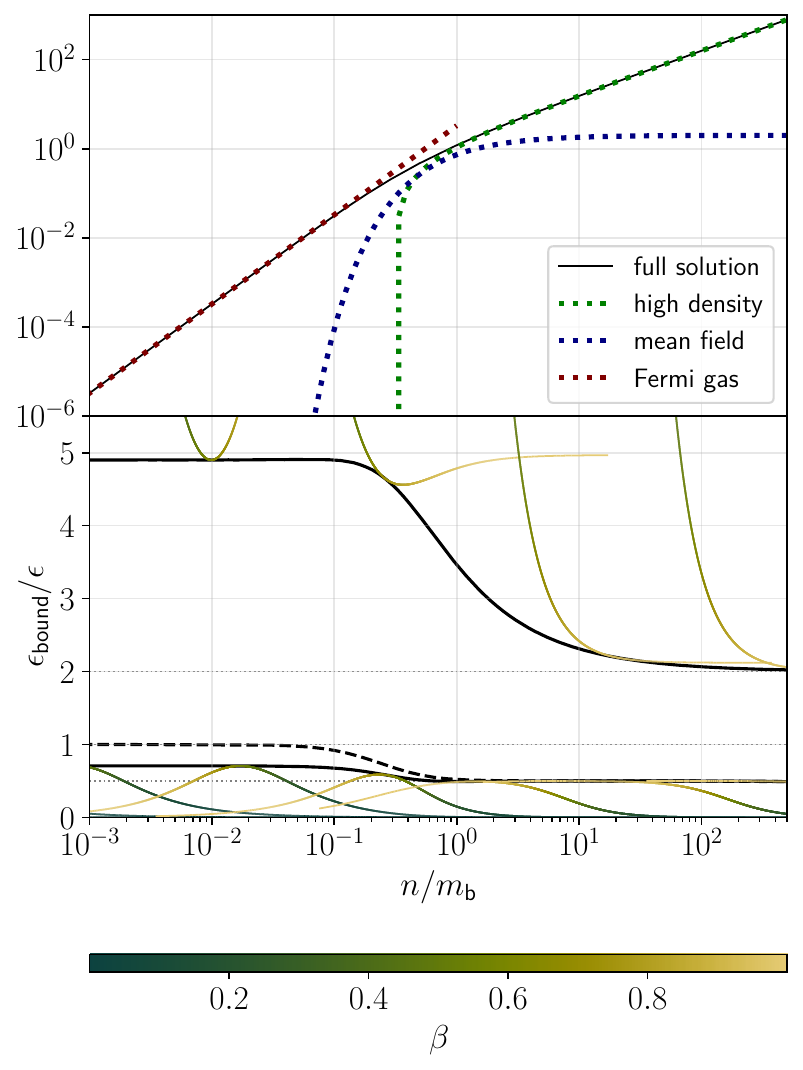}%
          }
          \caption{
                \textbf{Top}: The energy density $ \epsilon $, obtained by solving the integral equation \labelcref{eq:integral_eq_Haldane}, enters the interaction energy as a function of number density $ n $; the results are all in units of the boson mass $ m_\mathrm{b} $. 
                The regimes of high density (\cref{eq:high_desity_limit}), low density mean field (\cref{eq:mean_field_regime}) and very low density Fermi gas (\cref{eq:Fermi_gas_regime}) are indicated.
                \textbf{Bottom}: The ratio of upper and lower bounds (with the best lower bound shown in dashed lines) to the exact result. 
                A selection of curves corresponding to different by ${ \beta \in [10^{-6}, 0.9999] }$ ``boosted'' current densities are plotted with the bound being the envelope of all curves.
          }
          \label{fig:results_weak_strong_coupling}
        \end{figure*}

    Our numerical results for both the \gls{eos} at fixed number density and the bounds obtained using solutions at fixed current are shown in \cref{fig:results_weak_strong_coupling} for both weak and strong \gls{sgm} coupling.  
    As expected, the energy density as a function of density is well approximated at low densities by a Fermi gas and at high densities by the high density approximation of \cref{eq:high_desity_limit} for both strong and weak coupling.  Away from these two limits, the weak coupling theory is well described by mean-field theory (or equivalently via classical bosonic theory).  The results for the upper and lower bounds are of particular interest.
    The best lower bound uses the value of $ p / \epsilon $ obtained from the numerical solution to the finite number density integral equation. 
    Since the ratio becomes one at high densities, both lower bounds coincide in this regime.

\subsection*{Choice of current and boost for optimal bounds}

    First, consider the upper bound. 
    A low density $ n^\mathrm{u} $ is obtained by a small boost of some current or by some boost of a small current. 
    The latter provides a better bound because it keeps the boosted current $ j^\mathrm{u} $ low, bringing the system closer to the ground state with $ \epsilon ( n^\mathrm{u} ) $. 
    A high density results from either a large boost or a large current. 
    In this regime, the former provides a slightly better bound for the same reason as previously. 
    The curves become constant because an even larger boost keeps the ratio ${ n^\mathrm{u} / j^\mathrm{u} = \beta }$ nearly constant. 
    It must be as large as possible to approach the equality in \cref{eq:upper_bound}. 
    This explains why at low densities and with smaller boosts, the upper bound is worse than at large densities.
    
    Next, turn to the lower bounds before addressing the limiting values of ${ \epsilon_\mathrm{bound} / \epsilon }$. 
    A low density $ n^\mathrm{l} $ can lead to a moderate current from a large boost or to a small current from a moderate boost. 
    The latter provides a better bound because it keeps the density ${ n =  n^\mathrm{l} / \sqrt{ 1 - \beta^2 } }$ low in the boosted system, making the equality in \cref{eq:upper_bound} closer. 
    A high density can lead to some moderate current from a small boost or to a large current from a moderate boost. 
    The latter provides a better bound.

\subsection*{Low densities}

    At very low densities, where the Fermi gas description is valid, the ratio for the optimal upper bound approaches a value of about $ 4.90 $. 
    This can be understood in the following way:
   The system consists of nearly non-interacting particles. 
    The energy density is given by ${ \epsilon( n^\mathrm{u} ) = m_\mathrm{s} \, n^\mathrm{u} }$, while the pressure vanishes. 
    At zero number and non-zero current density $ j $ we compute
        \begin{align}
            & T_{tt} ( j ) = 2 \, m_\mathrm{s} \, j \, , 
            && T_{xx} ( j ) = m_\mathrm{s} \, j \, , 
        \end{align}
    using \cref{eq:final_density_of_states_integral_equation_kappa,eq:energy_momentum_tensor,eq:n_and_momentum}, ${ \rho ( \alpha ) \simeq \cosh( \alpha )/( 2\pi) }$, and expanding around the minimum $ \alpha_\mathrm{min} $, \cf{} \cref{eq:alpha_min}. 
    By \cref{eq:upper_bound,eq:def_densities_l_and_u}, the ratio
        \begin{align}
            \frac{ ( T_{tt} + \beta^2 \, T_{xx} ) / ( 1 - \beta^2 ) }{ \epsilon( n^\mathrm{u} ) } 
            = \frac{ ( 2 + \beta^2 ) / ( 1 - \beta^2 ) }{ \beta }
        \end{align}
    has a minimum of $ 2\sqrt{6} \approx 4.90 $ at $ \beta = \sqrt{2/5} \approx 0.63 $. 
    
    The best lower bound approaches the actual energy density. 
    This happens for a boost where ${ n^\mathrm{l} = j }$ with
        \begin{align}\label{eq:lower_bound_equality_condition}
            & \beta = 1 / \sqrt{2} \approx 0.7
            && \Leftrightarrow && \beta \, \gamma \equiv \beta / \sqrt{1-\beta^2} = 1 \, , 
        \end{align}
    and can be explained as follows: 
    The energy density is given by ${ \epsilon( n^\mathrm{l} ) = m_\mathrm{s} \, n^\mathrm{l} }$. 
    In the boosted frame, we find ${ T_{tt} = \gamma^2 \, \epsilon( n^\mathrm{l} ) }$. 
    Moreover, the energy density for the zero number but non-zero current density was computed to be ${ \epsilon(j) = 2 \, m_\mathrm{s} \, j}$. 
    The inequality \labelcref{eq:lower_bound} is derived from $ \epsilon(j) \leq T_{tt} $, \cf{} \cref{eq:derivation_lower_bound}. 
    We therefore conclude
        \begin{align}
            2 \, m_\mathrm{s} \, \beta \, \gamma \, n^\mathrm{l} \leq \gamma^2 \, m_\mathrm{s} \, n^\mathrm{l} \, .
        \end{align}
    The equality holds for the condition in \cref{eq:lower_bound_equality_condition}.

\subsection*{High densities}

    The upper and lower bounds constrain the energy density by a factor of two at high densities. 
    This can be understood in the following way: 
    The energy density grows quadratically with the number density, \cf{} \cref{eq:high_desity_limit}. 
    Besides, the energy density as a function of current density is the same function as for number density in this regime, 
        \begin{align}
            T_{tt} ( j ) = \epsilon(n)\vert_{n=j} \, .
        \end{align}
    The ways of filling up levels lead to the same function:
        \begin{itemize}[nosep]
            \item number density: filling all particle states from minus to plus the Fermi rapidity being very large;
            \item current density: filling up anti- and particle states from zero to a large rapidity values, \cf{} \cref{fig:states_filling}. 
        \end{itemize}
    For the upper bound one finds from \cref{eq:upper_bound,eq:def_densities_l_and_u}
        \begin{align}
            \epsilon ( n^\mathrm{u} ) \leq \frac{T_{tt} + \beta^2 \, T_{xx}}{\beta^2} \, \Big( \frac{n^\mathrm{u}}{j} \Big)^2
            \xrightarrow{\beta\rightarrow 1} 2 \, T_{tt} \, \Big( \frac{n^\mathrm{u}}{j} \Big)^2 \, , 
        \end{align}
    where ${ T_{tt} = T_{xx} }$ holds at high densities.
 This relation and \cref{eq:lower_bound,eq:def_densities_l_and_u} are utilized for the lower bound to express
        \begin{align}
            \epsilon ( n^\mathrm{l} ) \geq \frac{ \beta^2 \, T_{tt} }{ 1 + \beta^2 } \, \Big( \frac{n^\mathrm{l}}{j} \Big)^2
            \xrightarrow{\beta\rightarrow 1} \frac{ T_{tt} }{ 2 } \, \Big( \frac{n^\mathrm{l}}{j} \Big)^2 \, . 
        \end{align}
   Note that this understanding of the high density limits is valid for the ${ (1 + 1) }$-dimensional model and would need to be modified for higher dimensions.

\acknowledgments

    EO gratefully acknowledges the support of the Stiftung der Deutschen Wirtschaft (sdw) and thanks the University of Maryland (UMD) for its hospitality. TDC was supported in part by the U.S. Department of Energy, Office of Nuclear Physics under Award Number(s) DE-SC0021143, and DE-FG02-93ER40762.


\appendix
\section{Non-zero current density and zero number density ground state occupation}
\label{app:current_dens_ground_state}

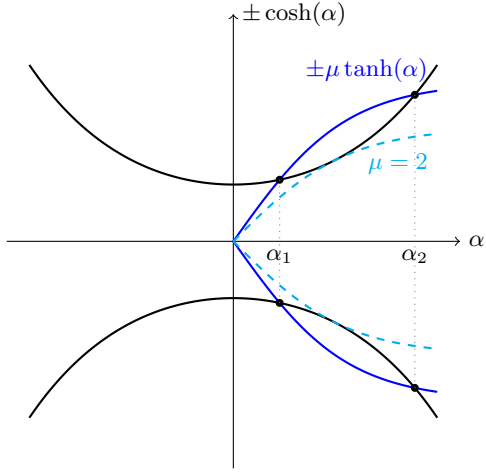
\begin{figure}[htbp]
            \centering
            \begin{tikzpicture}[scale=1.5]
  \def\Mu{2.8}
  \draw[->] (-2., 0) -- (2., 0) node[right] {$\alpha$};
  \draw[->] (0, -2.) -- (0, 2.) node[right] {$\pm\cosh(\alpha)$};
    
  \draw[black, thick, domain=-1.8:1.8, samples=100] 
    plot (\x, {0.5*cosh(\x)});
  \draw[blue, thick, domain=0:1.8, samples=100] 
    plot (\x, {0.5*\Mu*tanh(\x)}) node[above left] {$\pm\mu \tanh(\alpha)$};
  \draw[cyan, thick, dashed, domain=0:1.8, samples=100] 
    plot (\x, {0.5*2*tanh(\x)}) node[below left, yshift=-4pt] {$\mu = 2$};
 
  \draw[black, thick, domain=-1.8:1.8, samples=100] 
    plot (\x, {-0.5*cosh(\x)});
  \draw[blue, thick, domain=0:1.8, samples=100] 
    plot (\x, {-0.5*\Mu*tanh(\x)});
  \draw[cyan, thick, dashed, domain=0:1.8, samples=100] 
    plot (\x, {-0.5*2*tanh(\x)});
  
  \fill[black] (0.408728, {0.5*cosh(0.408728)}) circle (1pt);
  \fill[black] (1.60164, {0.5*cosh(1.60164)}) circle (1pt);
  \fill[black] (0.408728, {-0.5*cosh(0.408728)}) circle (1pt);
  \fill[black] (1.60164, {-0.5*cosh(1.60164)}) circle (1pt);
  
  \draw[gray, thin, dotted] (0.408728, {0.5*cosh(0.408728)}) -- (0.408728, 0);
  \draw[gray, thin, dotted] (1.60164, {0.5*cosh(1.60164)}) -- (1.60164, 0);
  \draw[gray, thin, dotted] (0.408728, {-0.5*cosh(0.408728)}) -- (0.408728, 0);
  \draw[gray, thin, dotted] (1.60164, {-0.5*cosh(1.60164)}) -- (1.60164, 0);
  
  \node[below] at (0.408728, 0) {$\alpha_1$};
  \node[below] at (1.60164, 0) {$\alpha_2$};
\end{tikzpicture}
            \caption{
                States of rapidities ${ \alpha_1 \leq \alpha \leq \alpha_2 }$ are populated in the non-zero current ground state. 
                All quantities are in units of $ m_\mathrm{s} $; 
                the energy per state of rapidity $ \alpha $ is drawn in black, its current contribution times the Lagrange multiplier $ \mu $ in blue. 
            }
            \label{fig:states_filling}
        \end{figure}

    In this appendix, we outline how to populate the available states such that we get the ground state at non-zero current density. 
    The vacuum state of zero number and current density would correspond to filling all ``negative energy'' states in the language of the Dirac sea. 
    A non-zero density at zero temperature is obtained by filling (or unfilling) all states up to the Fermi rapidity. 
    Here, a non-zero current but zero number density is achieved by simultaneously populating positive and negative energy states of equal spatial momentum such that their velocities are opposite, \ie 
        \begin{align}\label{eq:complex_rapidity}
            & \alpha \leftrightarrow \ii \pi - \alpha \, , 
            & \frac{p^{\mu}}{m_\mathrm{s}} = \begin{pmatrix}
                \cosh{ \alpha } \\ \sinh{ \alpha }
            \end{pmatrix} \leftrightarrow \begin{pmatrix}
                -\cosh{ \alpha } \\ \sinh{ \alpha }
            \end{pmatrix} \, . 
        \end{align}
    Furthermore, we must minimize
        \begin{align}
            \epsilon - \mu \, j \, ,
        \end{align}
    where $ \mu $ is the Lagrange multiplier fixing a non-zero current density.
    This amounts to filling the states around the minimum of $\epsilon ( \alpha) / j ( \alpha) $ up to $ \mu $, which is illustrated in \cref{fig:states_filling}. 
    We find that the ratio
        \begin{align}
            \frac{\epsilon ( \alpha )}{j ( \alpha)} = \, & m_\mathrm{s} \, \frac{ \cosh(\alpha)}{ \tanh(\alpha)}
        \end{align}
    has a minimum of $ 2 m_\mathrm{s} $ at $ \alpha_\mathrm{min} $,
        \begin{align}\label{eq:alpha_min}
            \alpha_\mathrm{min} = \ln( \sqrt{2} + 1 ) \, .
        \end{align}
    Hence, we require $ \mu \geq 2 \, m_\mathrm{s} $. 
    The recipe is as follows:
        \begin{enumerate}[nosep]
            \item Fix a value for $ \mu / m_\mathrm{s} \geq 2 $;
            \item Find the rapidities such that ${ \epsilon ( \alpha_{1,2} ) / j ( \alpha_{1,2} ) = \mu }$;
            \item Solve the integral equation \labelcref{eq:final_density_of_states_integral_equation_kappa} for $ \rho ( \alpha ) $;
            \item Calculate $ T^{\mu\nu} $ and $ j $ by \cref{eq:j_and_T}.
        \end{enumerate}
    We note that we assume a positive current density $ j > 0 $ and positive rapidities $ \alpha_{1,2} > 0 $ without loss of generality.

\section{Bethe ansatz and integral equation for density of states at finite current density}
\label{app:Bethe_an_int_eq}

    In the following, we present the derivation of the integral  equation \labelcref{eq:final_density_of_states_integral_equation_kappa}. 
    It is based on a Bethe ansatz that Bergknoff and Thacker used to study the vacuum properties of the model \cite{Bergknoff:1978bm,Thacker:1980ei}. 
    The general idea of the Bethe ansatz for a relativistic, fermionic system is to construct its ground state by starting from a nonphysical reference state -- conveniently, the empty Dirac sea. 
    Putting the theory in a box, one fills up the state levels successively, taking into account scattering with all previously added particles. 
    For local scatterings in one dimension, a phase shift variable captures all scattering effects. 
    This construction results in an integral equation for the density of states ${ \rho ( \alpha ) }$ in the infinite volume limit. 
    The density of states appears on both sides of the equation because filling a high-lying state influences all lower-lying levels as well. 
    Filling the entire Dirac sea corresponds to the vacuum theory. 
    Haldane unfills the sea again to derive the finite density result \cite{Haldane1982}. 
    We follow these ideas, but we must (un)fill the levels beyond the vacuum as described in \App\labelcref{app:current_dens_ground_state}. 
    
    First, we again point out the map from \cref{eq:complex_rapidity}, which converts a particle into a hole (and vice versa) with the same spatial momentum. 
    Therefore, we assume that the rapidity variable $ \beta $ is complex with an imaginary part of $ 0 $ or ${ \pm \ii \pi }$. 
    The local interaction between two particles/holes of the \gls{mtm} can be characterized by their phase shift,
        \begin{align}
            & \phi ( \xi ) 
            = -\ii \, \ln \left(
                -\frac{
                    \sinh \left[ \frac{1}{2} \, ( \xi - 2 \, \mu ) \right]
                }{  
                    \sinh \left[ \frac{1}{2} \, ( \xi + 2 \, \mu ) \right]
                }
            \right) \, ,
            & \xi \equiv \Delta \beta \, ,
        \end{align}
    where $ \mu $ is related to the coupling constant as follows
        \begin{align}
            \cot \mu = \, & -\frac{g}{2} \, . 
        \end{align}
    We list the derivative of the phase shift ${ K ( \xi ) \equiv \phi' ( \xi ) }$, ${ \xi \in \mathbb{R} }$, and its Fourier transform \cite[\Eq(2.82)]{Thacker:1980ei}
        \begin{subequations}\label{eq:kernel_K}
            \begin{align}
                K ( \xi )
                = \, & \frac{ \sin{ 2 \mu }}{ \cosh{ \xi } - \cos{ 2 \mu } } \, , 
                \\
                \bar{K} ( y ) \equiv \frac{ \tilde{K} ( y ) }{2 \pi} = \, & 
                \frac{
                    \sinh[ (\pi - 2 \mu) \, y ]
                }{ 
                    \sinh{ \pi y }
                } \, , 
            \end{align}
        \end{subequations}
    as well its counterpart ${ L ( \xi ) \equiv K ( \xi \pm \ii \pi ) }$ for later use,
         \begin{align}
            L ( \xi ) 
            = \, & \frac{ \sin{2 \mu} }{ - \cosh{ \xi } - \cos{ 2 \mu } } \, ,
            \\[8pt]
                \bar{L} ( y ) \equiv \frac{ \tilde{L} ( y ) }{2 \pi} = \, & 
                    \begin{cases}
                        \dfrac{
                            \sinh( - 2 \, \mu \, y )
                        }{ 
                            \sinh( \pi \, y ) 
                        }, 
                        & \mu < \dfrac{\pi}{2} \, , 
                        \\[12pt]
                        \dfrac{
                            \sinh[ 2 \, (\pi - \mu) \, y ]
                        }{ 
                            \sinh( \pi \, y ) 
                        }, 
                        & \dfrac{\pi}{2} < \mu < \pi \, .
                    \end{cases}
        \end{align}
    \textit{Remark:} The Kernel $ L ( \xi ) $ also acquires a $ \delta ( \xi ) $-term because the phase shift $ \phi ( \xi ) $ has a jump of $ 2\pi $ at $ \mathrm{Re}\,\xi = 0 $ for $ \mathrm{Im}\,\xi = \pi $ \cite[\Eq(2.91b)]{Thacker:1980ei}. 
    We drop this term because the limits of integration do not include zero in the finite current density case. 

    Now follows the main derivation.
    In order to count states, we place the theory in a box of length $ L $ with periodic boundary conditions. 
    This gives rise to a set of equations to be satisfied by the state $ \vert \beta_1, \ldots, \beta_n \rangle $ to be an admissible state 
        \begin{align}\label{eq:bethe_ansatz_equations}
            M_0 L \sinh(\beta_i) = - 2 \pi n_i - \sum_{j \neq i}^n \phi( \beta_i - \beta_j ) \, .
        \end{align}
    The integers $ n_i $ label the branches of the logarithm of the Bethe ansatz equations \cite[\Eq(3.3)]{Bergknoff:1978bm}. 
    According to the Pauli principle, different states must be labeled by different integers, and the ground state is obtained by choosing $ n_{i+1} - n_i = 1 $ \cite[\Eq(2.26)]{Thacker:1980ei}. 
    Equation \labelcref{eq:bethe_ansatz_equations} holds for $ \mathrm{Im}(\beta_i) = \pi $.
    Next, we take the difference between neighboring states and introduce the density of states $ \rho(\beta) $ as 
        \begin{align}
            \rho(\beta) = \lim_{\Delta \beta \to 0, ( L M_0 ) \to \infty}\frac{1}{L \, M_0 \, \Delta \beta} \, ,
        \end{align}
    resulting in
        \begin{align}
            \cosh(\beta) = - 2 \pi \rho(\beta) - \int_{\ii \pi - \infty}^{\ii \pi + \infty} \dd \beta' \, K(\beta - \beta') \rho(\beta') \, ,
        \end{align}
    with the derivative of the phase shift ${ K(\beta) = \phi'(\beta) }$ given in \cref{eq:kernel_K}, and the limits of integration corresponding to filling up the entire Dirac sea \cite[\Eq(2.62a)]{Thacker:1980ei}. 
    This is the integral equation for the vacuum density of states which we denote as $ \tilde{\rho}_0(\beta) $. 
    The non-zero current density of states is labeled by $ \tilde{\rho}(\beta) $, and the difference by ${ \Delta \tilde{\rho}(\beta) \coloneq \tilde{\rho}(\beta) - \tilde{\rho}_0(\beta) }$. 
    We remark again that $ \beta $ has an imaginary part of $ \pi $. 
    The variable $ \alpha $ is used for real rapidities in the following. 
    Emptying states from the Dirac sea and adding states of the same momentum above the Dirac sea leads to a finite current density, 
        \begin{align}
            2 \pi \Delta \tilde{\rho}(\beta) 
            = \, & - \int_{\ii \pi - \infty}^{\ii \pi + \infty} \dd \beta' \, K(\beta - \beta') \Delta \tilde{\rho}(\beta') \notag
            \\
            & + \int_{\ii \pi + \alpha_1^h}^{\ii \pi + \alpha_2^h} \dd \beta' \, K(\beta - \beta') \tilde{\rho}(\beta') \notag
            \\
            & - \int_{\alpha_1^p}^{\alpha_2^p} \dd \alpha' \, K(\beta - \alpha') \tilde{\rho}(\alpha') \, .
        \end{align}
    We specify the limits of integration below. 
    It is useful to shift the variable ${ \alpha \coloneq \beta - \ii \pi }$, ${ \rho (\alpha) \coloneq \tilde{\rho}(\ii\pi + \alpha) }$ etc.,
        \begin{align}\label{eq:intermediate_density_of_states_integral_equation}
            & 2 \pi \Delta \rho (\alpha) 
            = - \int_{-\infty}^{\infty} \dd \alpha' \, K(\alpha - \alpha') \, \Delta \rho(\alpha') \notag
            \\
            & \quad + \int_{\alpha_1^h}^{\alpha_2^h} \dd \alpha' \, K(\alpha - \alpha') \, \rho(\alpha') \notag
            \\
            & \quad - \int_{\alpha_1^p}^{\alpha_2^p} \dd \alpha' \, K(\alpha - \alpha' + \ii \pi) \, \rho(\alpha'- \ii \pi) \, .
        \end{align}
    With this step, $ \rho (\alpha) $ (${ \alpha \in \mathbb{R} }$) still denotes the density of states for negative energy states. 
    If the third term was absent it would be the density of states for negative energy states with negative energy states unfilled just as in Haldane's work. 
    If the second term was absent it would be the density of states for negative energy states with positive energy states filled instead. 
    One would then have to analytically continue the result to arguments ${ \alpha + \ii \pi }$, hence positive energy states to obtain the equivalent equation. 

    After applying a Fourier transform to \cref{eq:intermediate_density_of_states_integral_equation}, bringing the first term of the right-hand side to the left-hand side, solving for $ \Delta \rho (\alpha) $ leads to the following solution after Fourier back transform 
        \begin{align}\label{eq:general_density_of_states_integral_equation}
            \rho (\alpha) = \, & \rho_0 (\alpha) + \rho_\mathrm{h} (\alpha) + \rho_\mathrm{p} (\alpha) \, ,
        \end{align}
    with the vacuum solution $ \rho_0 (\alpha) $ in \cref{eq:vacuum_density_of_states_solution} and 
    \begin{subequations}
        \begin{align}
            \rho_\mathrm{h} (\alpha) = \, &
            \int_{\alpha_1^h}^{\alpha_2^h} \dd \alpha' \int_{-\infty}^{\infty} \frac{\dd y}{2\pi} \, e^{\ii y (\alpha - \alpha')} \frac{ \bar{ K } ( y ) }{ 1 + \bar{ K } ( y )} \, \rho (\alpha') \, ,
            \\
            \rho_\mathrm{p} (\alpha) = \, &
            \int_{\alpha_1^p}^{\alpha_2^p} \dd \alpha' \int_{-\infty}^{\infty} \frac{\dd y}{2\pi} \, e^{\ii y (\alpha - \alpha')} \frac{ -\bar{ L } ( y ) }{ 1 + \bar{ K } ( y )} \, \rho (\alpha'-\ii\pi) \, .
        \end{align}
    \end{subequations}
    Before continuing, we recover the integral equation for non-zero number density by choosing ${ -\alpha_1^h = \alpha_\mathrm{F} = \alpha_2^h }$ and ${ 0 = \alpha_1^p = \alpha_2^p }$, ${ \alpha_\mathrm{F} > 0 }$ the Fermi rapidity,  \cite[\Eq(3.9)]{Haldane1982}
        \begin{align}\label{eq:finite_number_density_integral_equation}
            \rho (\alpha) = \, & \rho_0 (\alpha) + \int_{-\alpha_\mathrm{F}}^{\alpha_\mathrm{F}} \dd \alpha' \rho (\alpha') \notag 
            \\
            & \times \int_{-\infty}^{\infty} \frac{\dd y}{2\pi} \, e^{\ii y (\alpha - \alpha')} \, 
                \frac{ \bar{ K } ( y ) }{ 1 + \bar{ K } ( y )} \, . 
        \end{align}
   The result for a finite current density is a straightforward extension of the above by choosing different limits,
        \begin{subequations}
        \begin{align}
            & \alpha_1^h = \alpha_1 \, , && \alpha_2^h = \alpha_2 \, , 
            \\
            & \alpha_1^p = -\alpha_1 \, , && \alpha_2^p = -\alpha_2 \, . 
        \end{align}
        \end{subequations}
    Equation \labelcref{eq:general_density_of_states_integral_equation} becomes (after ${\alpha' \to -\alpha'}$, ${y \to -y}$) 
        \begin{align}\label{eq:density_of_states_integral_equation}
            & \rho (\alpha) = \rho_0 (\alpha) +
            \\
            & + \int_{\alpha_1}^{\alpha_2} \dd \alpha' \int_{-\infty}^{\infty} \frac{\dd y}{2\pi} \, 
            \frac{ 
                e^{\ii y (\alpha - \alpha')} \, \bar{ K } ( y ) 
            }{ 
                1 + \bar{ K } ( y )
            } \, \rho (\alpha') \notag
            \\
            & + \int_{\alpha_1}^{\alpha_2} \dd \alpha' \int_{-\infty}^{\infty} \frac{\dd y}{2\pi} \, 
            \frac{ 
                e^{\ii y (-\alpha - \alpha')} \, \bar{ L } ( y ) 
            }{ 
                1 + \bar{ K } ( y )
            } \, \rho (-\alpha'-\ii\pi) \, . \notag
        \end{align}
    We find by analytic continuation that $\rho (\alpha) $ and ${ \bar{\rho} (\alpha) \coloneq -\rho(-\alpha - \ii \pi) }$ satisfy the same integral equation. 
    The continuation requires some care. 
    Simply replacing $ \alpha $ with a complex variable is problematic, since the integral would not necessarily converge. 
    We also need to replace ${ \bar{K} \leftrightarrow \bar{L} }$ in the numerators of \cref{eq:density_of_states_integral_equation}. 
    This can be seen from carefully inspecting the Fourier transformations going back to the definition of $ \rho_\mathrm{p} (\alpha) $,
        \begin{align}
            & \rho_\mathrm{p} (\alpha) =
            -\int_{-\infty}^{\infty} \frac{\dd y}{2\pi} \, \frac{ e^{\ii y \alpha} }{ 1 + \bar{ K } ( y )} 
            \int_{-\infty}^{\infty} \dd \tilde{\alpha} \,e^{-\ii y \tilde{\alpha} } \notag
            \\
            & \quad \times \int_{\alpha_1^p}^{\alpha_2^p} \dd \alpha' \, K(\tilde{\alpha} - \alpha' + \ii \pi) \, \rho(\alpha'- \ii \pi) \, .
        \end{align}
    Equation \labelcref{eq:density_of_states_integral_equation} can thus be written as
        \begin{align}\label{eq:final_density_of_states_integral_equation}
            \rho (\alpha) = \, & \rho_0 (\alpha) + 
            \int_{\alpha_1}^{\alpha_2} \dd \alpha' \int_{-\infty}^{\infty} \frac{\dd y}{2\pi} \, 
            \frac{ 
                e^{-\ii \alpha' y} 
            }{ 
                1 + \bar{ K } ( y )
            }\notag
            \\
            & \times \left[
                e^{\ii y \alpha} \, \bar{ K } ( y ) - e^{-\ii y \alpha} \, \bar{ L } ( y )
            \right] \, \rho (\alpha') \, .
        \end{align}
    This is our central result. 
    We express it in a more convenient form in the remainder of this appendix. 
    The phase shift variable $ \mu $ can be expressed in terms of the variable $ \kappa $ introduced in \cref{eq:kappa_definition} as follows \cite[\Eq(4.18)]{Bergknoff:1978bm}
        \begin{align}
            \mu = \, & \frac{\pi}{ 1 + \kappa } \, .
        \end{align}
    The vacuum solution $ \rho_0 (\alpha) $ is given by
        \begin{align}\label{eq:vacuum_density_of_states_solution}
            \rho_0 (\alpha) = \rho_0 ( 0 ) \, \cosh \left( \tfrac{1 + \kappa}{2} \, \alpha \right) \, ,
        \end{align}
    where $ \rho_0 ( 0 ) $ is a diverging renormalization constant \cite[\Eq(3.11)]{Haldane1982} \& \cite[\Eq(4.2')]{Bergknoff:1978bm}. 
    Renormalizing the solution $ \rho $ and re-scaling ${ \alpha \, ( 1 + \kappa ) / 2 \to \alpha }$, ${ y \to y \, ( 1 + \kappa ) / 2 }$ leads to the final form of the integral equation \labelcref{eq:final_density_of_states_integral_equation_kappa} with kernels
    \begin{subequations}
        \begin{align}
            \bar{K} ( y ) = \, & 
                \frac{
                    \sinh[ \pi / 2 \, ( \kappa - 1 ) \, y  ]
                }{ 
                    \sinh[ \pi / 2 \, ( \kappa + 1 ) \, y  ] 
                } \, ,
            \\
            \bar{L} ( y ) = \, & 
                \begin{cases}
                    \dfrac{
                        \sinh[ -\pi \, y ]
                    }{ 
                        \sinh[ \pi / 2 \, ( \kappa + 1 ) \, y  ]
                    }, 
                    & 1 < \kappa< \infty \, , 
                    \\[12pt]
                    \dfrac{
                        \sinh[ \pi \, \kappa \, y  ]
                    }{ 
                        \sinh[ \pi / 2 \, ( \kappa + 1 ) \, y  ]
                    }, 
                    & 0 < \kappa< 1 \, .
                \end{cases}
        \end{align}
    \end{subequations}
    We can rewrite 
    \begin{subequations}
        \begin{align}
            \frac{ \bar{ K } ( y ) }{ 1 + \bar{ K } ( y ) } = \, & 
                \frac{1}{2} \, \bigg(
                    1 - \frac{ \tanh( \pi / 2 \, y ) }{ \tanh( \pi / 2 \, \kappa \, y ) }
                \bigg) \, ,
            \\[11pt]
            \frac{ \bar{ L } ( y ) }{ 1 + \bar{ K } ( y ) } = \, & 
                \begin{cases}
                    -\dfrac{
                        1 - \ee^{ -\pi y }
                    }{ 
                        1 - \ee^{ -\pi \kappa y }
                    } \, \ee^{ -\pi / 2 ( \kappa - 1 ) y } \, , 
                    & 1 < \kappa< \infty \, , 
                    \\[12pt]
                    \dfrac{
                        1 + \ee^{ -\pi \kappa y }
                    }{ 
                        1 + \ee^{ -\pi y }
                    } \, \ee^{ -\pi / 2 ( 1 - \kappa ) y } \, ,
                    & 0 < \kappa< 1 \, .
                \end{cases}
        \end{align}
    \end{subequations}

\section{Numerical setup}
\label{app:numerical_setup}

    The integral equations needed to be solved are second kind Fredholm equations. 
    We use the composite Newton--Cotes trapezoidal method to approximate the integral. 
    The number of grid points used is included in our uploaded data sets. 
    As the chemical potential / Lagrange multiplier increases, the integral becomes more important because the ``backflow effects'' become increasingly significant. 
    This requires more grid points than at low densities. 
    Our implementation follows \Reff\cite{Allhands:2022}. 

    The kernels of the integral equations contain integrals that we compute using the Scientific Python library's integrate package and quad function. 
    Importantly, we control the oscillatory behavior of the integrand using a weight of $ \cos $. 
    
    \vfill

\bibliography{ZZ_literature}

\end{document}